\documentclass[aps,prd,twocolumn,nofootinbib,superscriptaddress]{revtex4}
\usepackage[normalem]{ulem}
\usepackage{latexsym}
\usepackage{amssymb}
\usepackage{epsfig,amsmath,graphics}
\usepackage{pstricks}
\usepackage{color}
\usepackage{dcolumn}

\newcommand{\gsim}{\lower.7ex\hbox{$\;\stackrel{\textstyle>}{\sim}\;$}}
\newcommand{\lsim}{\lower.7ex\hbox{$\;\stackrel{\textstyle<}{\sim}\;$}}
\newcommand{\OO}{\mathcal{O}}

\newcommand{\gev}{\text{ GeV}}

\newcommand{\METrel}{\mbox{$E^{\text{rel}}_T\hspace{-0.25in}\not\hspace{0.18in}$}}
\newcommand{\MET}{\mbox{$E_T\hspace{-0.23in}\not\hspace{0.18in}$}}

\begin{document}

\pagestyle{plain}

\title{Electroweakinos Hiding in Higgs Searches }

\author{Mariangela Lisanti$^a$ and Neal Weiner$^{b}$\\
$^a$ PCTS, Princeton University, Princeton, NJ 08544\\
$^b$ Center for Cosmology and Particle Physics, \\Department of Physics, New York University, New York, NY 10003}

\begin{abstract}
Direct production of electroweakly charged states may not produce the high energy jets or the significant missing energy required in many new physics searches at the LHC.  However, because these states produce leptons, they are still potentially detectable over the sizeable Standard Model backgrounds.  We show that current LHC Higgs searches, particularly in the $WW^*$ and $ZZ^*$ channels, are sensitive to new electroweak states, such as supersymmetric charginos or neutralinos.  Indeed, the 1.7 fb$^{-1}$ Higgs searches can provide the strongest limits in certain regions of parameter space, extending the LEP bound up to $\sim200$ GeV in some cases.  Additionally, electroweakino production can form an interesting physics background for Higgs searches, especially at low luminosity and statistics.  We show that dilepton searches with low missing energy requirements are complementary to existing searches in exploring regions of parameter space where new electroweak states are light or have compressed spectra.
\end{abstract}

 \maketitle

\section{Introduction}

The hierarchy problem is the principle motivation for new physics at the weak scale, compelling us to search for new fields that regulate the quadratic divergences in the Higgs boson mass.  Many theories Beyond the Standard Model (BSM) posit the existence of new colored states at the weak scale to cancel the largest divergence, which arises from the top quark loop \cite{Martin:1997ns}.  The production rates at the LHC of these new states can be sizeable, even if their masses are large, and they are observable in the jets$+X$ channels.  

At the same time, the Higgs self-coupling and gauge loops should also be regulated, but likely by states with only electroweak quantum numbers.  These states are produced at a lower rate at the LHC, but are equally important in their implications for the weak scale.  Even outside of the hierarchy problem, one might make a simple argument for such new electroweak states. While chiral matter is light for a clear reason, the Higgs is light for an unknown reason. It may be that the weak scale is simply a special mass scale for weakly charged fields. Other, similar fields might exist, of which we would be unaware  if they did not acquire vevs or couple to Standard Model (SM) fields in any significant way. Only direct searches can answer the question as to whether such states exist.

The strongest model-independent bounds on light electroweak states come from LEP, which searched for chargino pair production and neutralino associated production in electron-positron collisions~\cite{Abbiendi:2003sc}.  Limits were set at the 95\% confidence level for $\sigma\times$Br in the hadronic, semi-leptonic, and fully leptonic channels.  The analysis at $\sqrt{s} =$ 192--209 GeV bounds the neutralino mass splitting over a wide range of parameter space, and sets a lower bound on the chargino mass of $\sim100$ GeV.

To date, the Tevatron and LHC have searched for electroweakinos in the decays of new colored particles~\cite{Aad:2011cw, ATLAS:2011iu, CMS-SUS-010,CMS-SUS-011,CMS-SUS-015} and in multilepton final states~\cite{Abazov:2009zi, CDFtrileptons, cmstalk, CMS-PAS-EXO-11-045, CMS-SUS-013,CDF-EXO-10464,DZero-5126-CONF}.  The former searches take advantage of the large production cross section of the new colored states (i.e., squarks or gluinos) to increase production of the electroweakinos.  These searches typically require several jets as well as large missing energy.  However, the interpretation of the results depends on the branching fractions into electroweakinos, which may be small depending on the spectrum of the theory.  A recent ATLAS analysis for dileptons~\cite{Aad:2011cw} looked at channels both with and without jet activity.  The same-sign lepton channel with no jet activity was particularly sensitive to direct electroweakino production, but did not probe compressed spectra because of a large missing energy requirement of 100 GeV.     
\begin{figure}[t] 
\includegraphics[width=0.45\textwidth,angle=0]{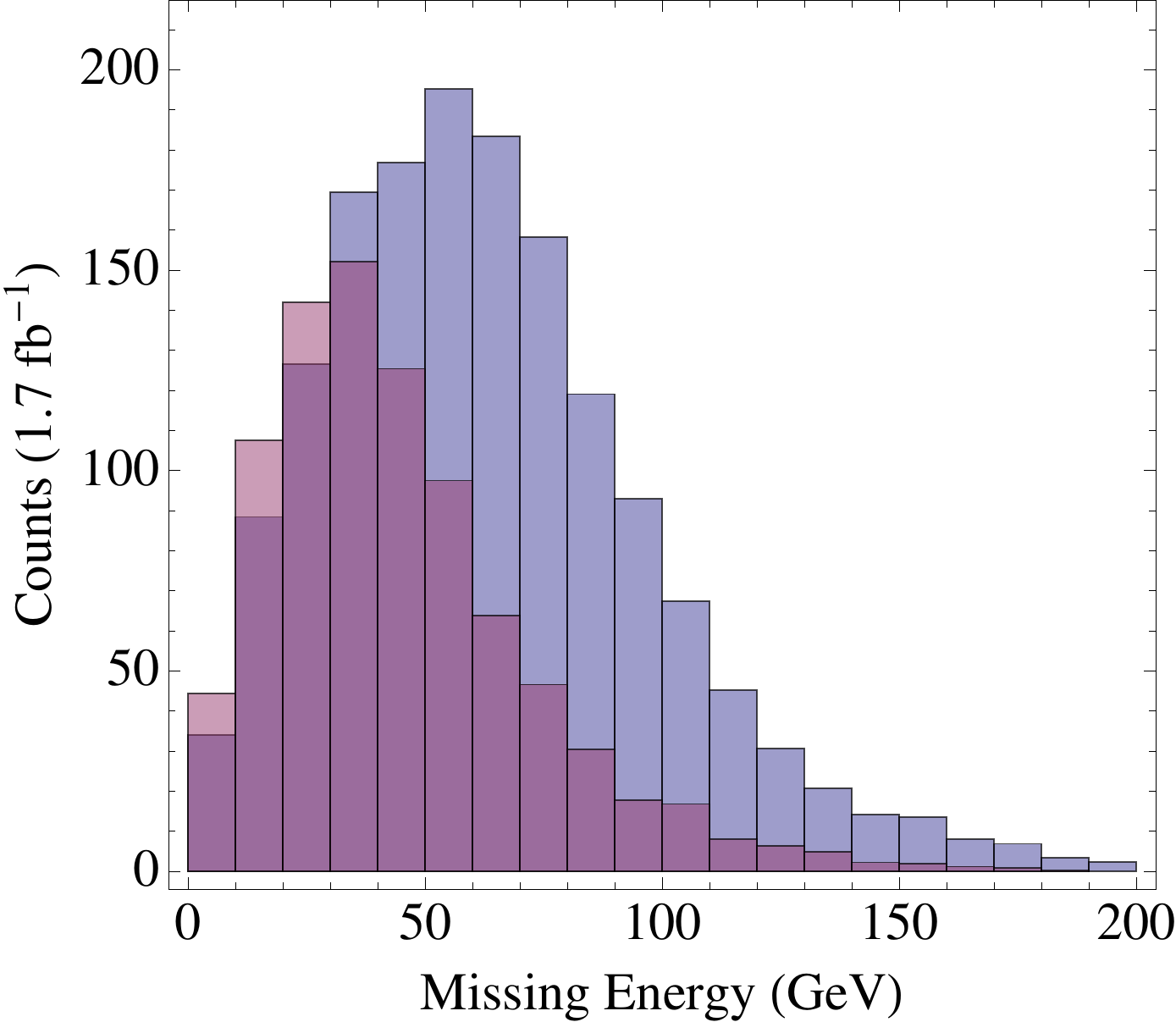}
   \caption{Missing energy distribution for associated production of 120 GeV Winos decaying to a 1 GeV (blue) or 80 GeV (pink) Bino.  Events satisfy the base selection cuts for the ATLAS $WW^*$ search~\cite{ATLAS-CONF-2011-134}, before application of the $\METrel$ cut. }
   \label{fig:methist}
\end{figure}

As an alternative, one can focus on more lepton-rich final states.  Trileptons are a quintessential signature of SUSY~\cite{Nath:1987sw,Barbieri:1991vk,Baer:1992dc,Baer:1993tr,Baer:1994nr,Lopez:1994dm,Forrest:2009gm}, arising from associated production of charginos and neutralinos that decay through electroweak gauge bosons.  (For a recent review, see~\cite{Dube:2008zz}.)  The CDF trilepton analysis with 5.8 fb$^{-1}$ sets a 95\% confidence level lower limit of 168 GeV on an mSUGRA chargino~\cite{CDFtrileptons}.  Model-independent trilepton searches have been suggested~\cite{Dube:2008kf}, especially in the context of simplified models~\cite{Alves:2011wf}.  The CMS Collaboration currently has generic searches for multileptons underway~\cite{CMS-SUS-013,cmstalk}, with specific interpretations in the context of e.g., leptonic R-parity violation \cite{CMS-PAS-EXO-11-045}. 

While there are many SUSY searches for light states unaccompanied by hard hadronic elements, most require a significant amount of missing energy.  However, if the mass spectrum for the electroweakinos is compressed, many events can have missing energy below the typical $100 \gev$ cut.  This is illustrated in Fig.~\ref{fig:methist}, which shows the distribution of missing energy for a $120 \gev$ Wino decaying to a $1 \gev$ (blue) or $80 \gev$ (pink) Bino.  Clearly, for certain parameter choices, the missing energy can peak well below 100\gev, and in certain cases even below 50\gev, making fairly inclusive approaches important in searching for these particles.  Fortunately, such searches are already underway at the LHC---namely, the Higgs searches.  In particular, Higgs searches that require two and four-lepton final states (i.e., from $WW^*$ and $ZZ^*$) generally have much weaker cuts on missing energy and are therefore sensitive to a broad range of parameter space for new electroweak states. 

\begin{figure}[b] 
\includegraphics[width=0.45\textwidth,angle=0]{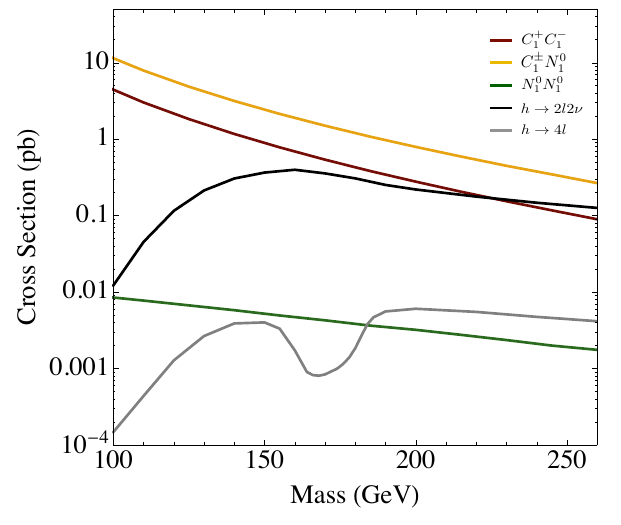}
\caption{NLO $\sigma \times {\rm BR}$ calculations for two and  four-lepton Higgs modes, as well as electroweakino production at the LHC.  $C^\pm_1 N^0_1$  (chargino - neutralino) and $C^\pm_1 C^\mp_1$  production arise from s-channel vector bosons,  while $N_1^0 N_1^0$ is generated from the t-channel exchange of $800 \gev$ squarks.}
\label{fig:xsection}
\end{figure}

Figure~\ref{fig:xsection} compares the NLO production cross sections for electroweakinos, obtained using~\texttt{Prospino 2.0}~\cite{Beenakker:1999xh}, with those for two and four-lepton Higgs modes~\cite{LHCHiggsCrossSectionWorkingGroup:2011ti}.  
The cross section for chargino pair (red) and neutralino-chargino associated (yellow) production is as large as, or even larger than, that expected for a Higgs decaying to two leptons (black).  Similarly, the neutralino pair production cross section (green) compares favorably with that for the Higgs four-lepton decay mode (gray).  Therefore, direct electroweakino production has the appropriate cross section and final states to be relevant for Higgs searches. 

This Letter will explore two questions simultaneously.  First, to what extent do present Higgs searches constrain light, weakly charged states?  Second, can such states be a physics background in Higgs searches, potentially even being confused for a Higgs at low statistics?  We discuss these questions in the context of a simplified model inspired by a light Bino and Wino in supersymmetry (Sec.~\ref{sec: Simplified Model}) and determine the constraints on light electroweakinos from current Higgs searches (Sec.~\ref{sec: Searches}).  Finally, we comment on modifications to the Higgs searches that could improve the discovery potential of a simple RECASTing~\cite{Cranmer:2010hk} of the limits (Sec.~\ref{sec: Conclusions}).

\section{A Simplified Model}
\label{sec: Simplified Model}
\begin{figure*}[t] 
\begin{center}
\includegraphics[width=\textwidth,angle=0]{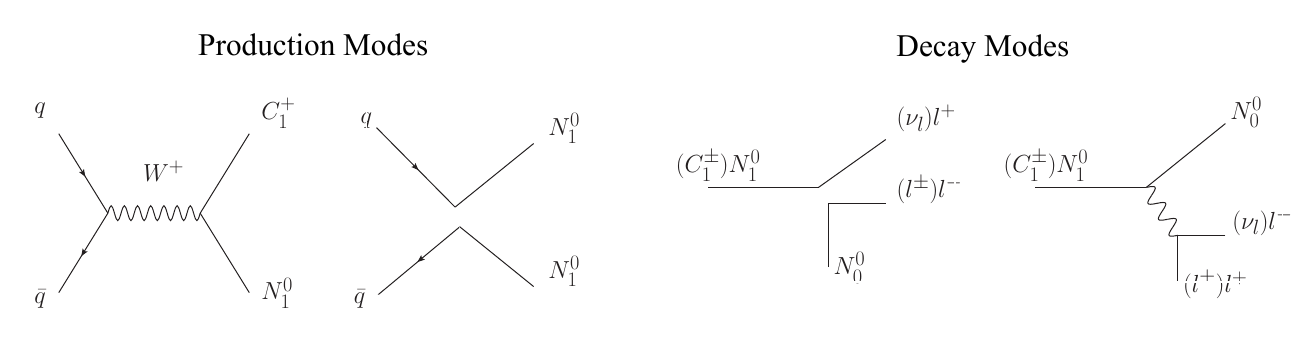}
\end{center}
   \caption{Production mechanisms and decay modes for the simplified model.  The $C_1^\pm C_1^\mp$ production mode is similar to that for $C_1^\pm N_1^0$, except with an s-channel $\gamma/Z^0$ boson.}
   \label{fig:simproduce}
\end{figure*}

This section introduces the simplified model that we will use to illustrate the sensitivity of Higgs searches to light electroweak states.  This simplified model is motivated by the supersymmetric scenario of electroweakinos that decay via off-shell sleptons or gauge bosons. 
It includes a complete SM singlet $N_0^0$ with mass $M_0$ as well as an $SU(2)$-triplet $C^\pm_1, \, N_1^0$ at mass $M_1$. We consider $C_1^\pm N_1^0$ production through an s-channel $W^\pm$, $C_1^\pm C_1^\mp$ pair production through an s-channel $\gamma/Z^0$,  as well as $N_1^0 N_1^0$ pair production through a higher dimension operator (in SUSY, via a t-channel squark), as shown in Fig.~\ref{fig:simproduce}.

Because we are interested in the production of {\em light} particles ($\sim 100$--$200 \gev$), the jets and missing energy are relatively weak and lepton triggers are necessary.  As illustrated in Fig.~\ref{fig:simproduce}, the $C_1^{\pm}$ and $N_1^0$ can decay into $N_0^0$ and leptons either through a $W^\pm$/$Z^0$ (arising via Higgsino mixing in SUSY) or a higher dimension operator (arising from an off-shell slepton).  As a consequence, this simplified model has 2-, 3- and 4-leptons, with the 4-lepton signal relying upon a pair production mechanism for $N_1^0 N_1^0$.  
 \hskip 0.1in
\begin{figure}[b] 
\includegraphics[width=0.45\textwidth,angle=0]{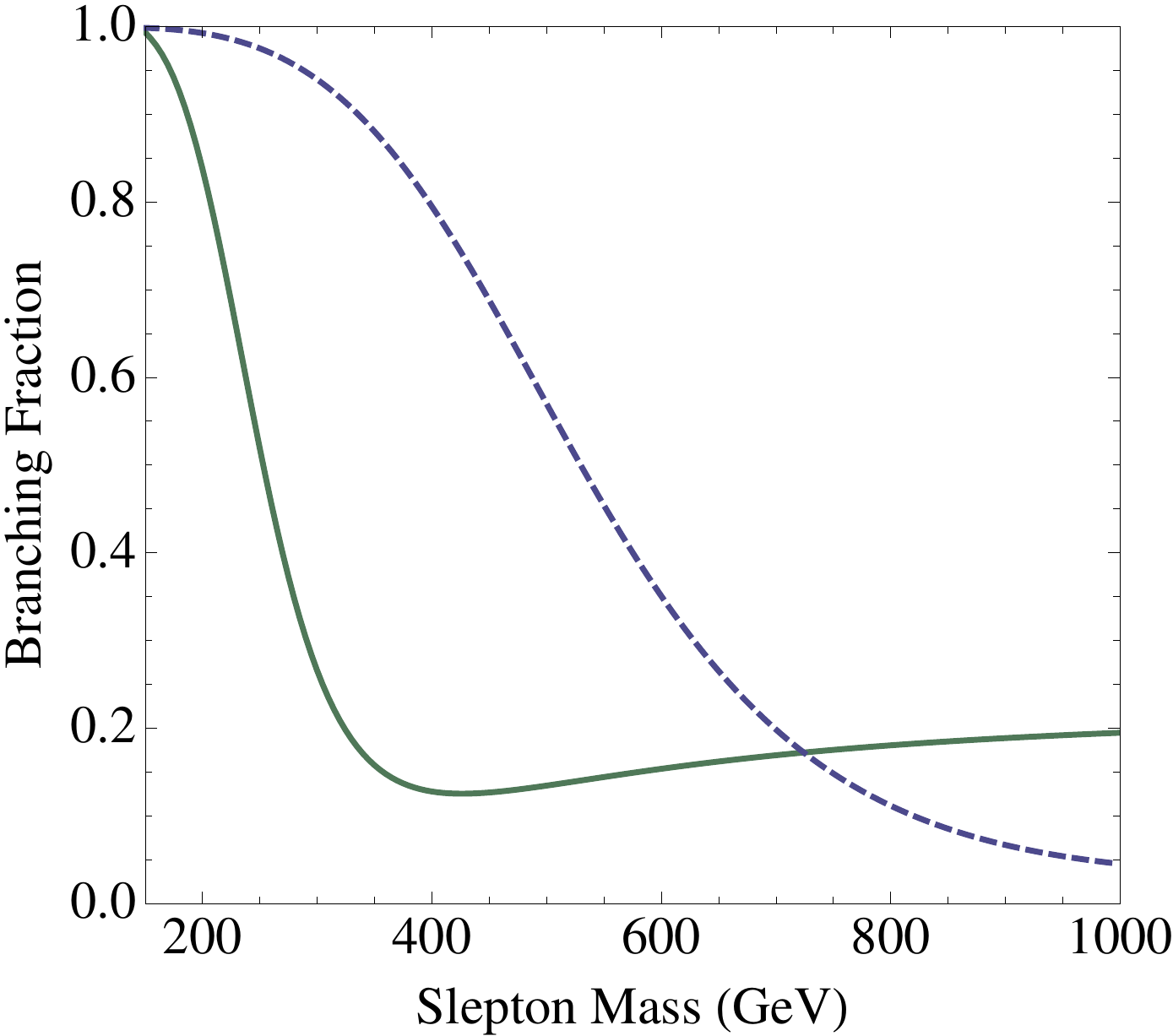}
   \caption{Branching ratio of $N^0_1\rightarrow l^+ l^- N_0^0$ (blue dashed) and $C_1^\pm\rightarrow l^{\pm} \nu N_0^0$ (green) for $l = e, \mu$ as a function of the slepton mass.  In this example, $M_1 = 150 \gev$, $M_0 = 100 \gev$, and the Higgsino has a mass of 1 TeV.}
   \label{fig:susyBRs}
\end{figure}

Before presenting the detection prospects of this simplified model in Higgs searches, let us consider how reasonable its phenomenology is in the framework of supersymmetry.  Specifically, how often do leptonic decays dominate the final states of electroweakinos?  Consider the supersymmetric version of this simplified model, augmented with sleptons and Higgsinos.  Figure~\ref{fig:susyBRs} shows the branching ratio of $C_1^{\pm} \rightarrow N_0^0 l^\pm \nu_l$ and $N_1^0 \rightarrow N_0^0 l^+ l^-$, for $(M_0, M_1) = (100, 150) \gev$ and 1 TeV Higgsinos, as calculated by Susy-Hit~\cite{Djouadi:2006bz}.  There is a sizeable leptonic branching ratio for $N_1^0$ (blue line) for a fairly broad range of slepton masses.  In contrast, the leptonic branching ratio for $C_1^\pm$ (green line) rises above the level for pure $W^\pm$ boson decays only for the lightest sleptons.  Because the $W^\pm$ boson already produces a sizeable number of leptons (compared especially to the $Z^0$), interesting limits are present so long as the leptonic branching ratio of $N_1^0$ is not too small. Nonetheless, the simplified models we choose are close to what is realized in SUSY models.

Moreover, it is worth noting here that {\em both} decays shown in Fig.~\ref{fig:simproduce} arise from heavy physics. The decay through the off-shell slepton is clearly dependent on the mass of the slepton, and the operator that allows decays through $W$'s and $Z$'s arises from integrating out Higgsino states, i.e., there is no natural decay of $N_1^0,C_1^\pm \rightarrow N_0^0$ without higher dimension operators. Consequently, either decay mode (leptonic or Z/W-dominated) could completely dominate naturally in other models of new physics.

\section{Electroweakino Limits from Higgs Searches}
\label{sec: Searches}

\subsection{Event Generation and Selection}
Both CMS and ATLAS have Higgs searches for  $WW^*$~\cite{ATLAS-CONF-2011-134,CMS-HIG-11-024} and $ZZ^*$~\cite{ATLAS-CONF-2011-162,CMS-HIG-11-025} final states.  In this section, we will take the ATLAS cuts as the primary example.

We consider the $h\rightarrow WW^*$ search that requires two isolated opposite-sign leptons~\cite{ATLAS-CONF-2011-134}.  The leading lepton must have $p_T > 25 \gev$ and the sub-leading lepton must have $p_T > 20 \text{ (15)} \gev$ for electrons (muons).  For same flavor events, $m_{ll}>15 \gev$ and $|m_Z-m_{ll}|>15 \gev$ is additionally required, while for opposite flavor events, $m_{ll}>10 \gev$ is required.  For same (opposite) flavor events, $\METrel > 40 \text{ (25)} \gev$, where
\begin{eqnarray}
\METrel = \begin{cases}
\MET& \text{ if $\Delta \phi \ge \pi/2$}\\
\MET\cdot \sin \Delta\phi & \text{ if $\Delta \phi < \pi/2$}
\end{cases} \,
\end{eqnarray}
and $\Delta \phi$ is the difference in the azimuthal angle between the missing energy vector and the nearest lepton or jet.  The $h\rightarrow WW^*$ analysis is divided into a 0-jet and 1-jet channel, where a jet must satisfy $p_T > 25 \gev$ and $|\eta| < 4.5$.  The 0-jet channel requires no jets and dileptons satisfying $p_T^{ll}> 30 \gev$ and $m_{ll}<50 \gev$.  The 1-jet channel consists of exactly one jet, with b-jets vetoed.  The event is required to be relatively soft, with the vector sum $| {\bf p}_T^{l_1} + {\bf p}_T^{l_2}+{\bf p}_T^{j}+{\bf p}_T^{\text{miss}}|< 30 \gev$.  Finally, to control $Z\rightarrow \tau \tau$, events with reconstructed $\tau$'s with $|m_{\tau\tau} - m_Z|<25 \gev$ are excluded.  The Higgs search also has additional cuts on the dilepton opening angle and transverse mass, which we do not include here because they are optimized for a Higgs signal.  The dominant SM backgrounds are $WW$, $W+$jets, $Z/\gamma^*+$jets, $t\bar{t}$, $tWb/tb/tqb$, $WZ/ZZ/W\gamma$, which contribute a total of $139\pm20$ events in the 0-jet channel and $64\pm10$ events in the 1-jet channel at 1.7 fb$^{-1}$ (see Tables 4 and 8 in~\cite{ATLAS-CONF-2011-134}).
\hskip 0.1in
\begin{figure}[b] 
\includegraphics[width=0.45\textwidth,angle=0]{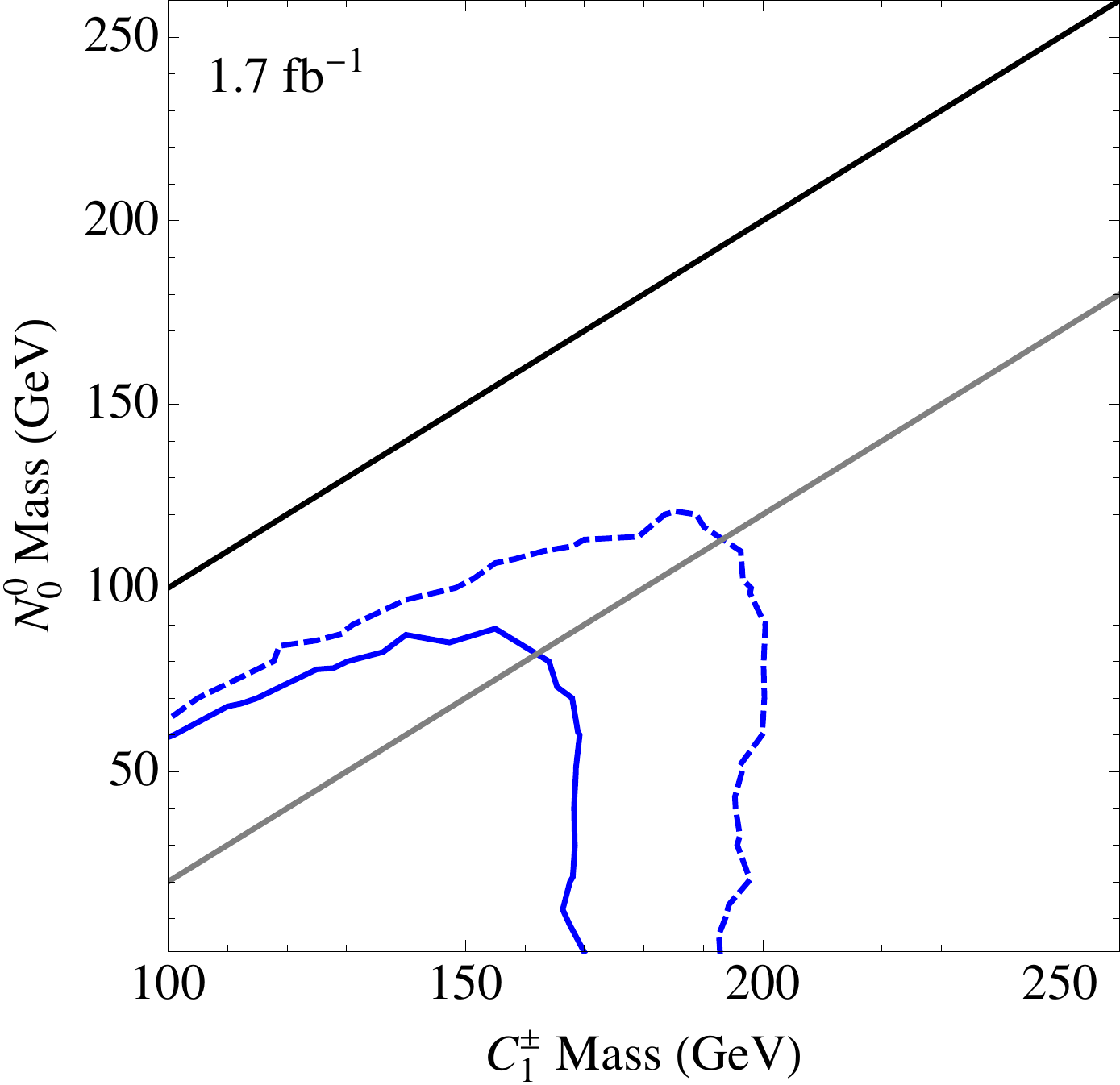}    \caption{Sensitivity of the ATLAS $2$-lepton, 0-jet Higgs search to $C_1^{\pm} C_1^{\mp}$ production with lepton-dominated decays.  Dashed and solid lines indicate the 1 and 2$\sigma$ sensitivities, respectively.  The region above the solid black line is forbidden by kinematics.  On-shell $W$-boson decays are kinematically allowed below the solid gray line; if not highly suppressed, these decays will dominate in this region.}
   \label{fig:charginos}
\end{figure}

A separate analysis is performed for four-lepton final states~\cite{ATLAS-CONF-2011-162}.  Events are triggered by an electron (muon) with $E_T \gtrsim 20$--$22$ ($p_T > 18$).  All events are required to have two pairs of same-flavor, opposite-sign isolated leptons, each with $p_T > 7 \gev$ and $|\eta| < 2.47 \text{ (2.7)}$ for electrons (muons).  At least two leptons are required to have $p_T > 20 \gev$.  All the leptons in the event must be separated by $\Delta R > 0.1$ and the lepton pair best reconstructing the $Z^0$ boson is required to have $|m_{Z} - m_{12}|< 15 \gev$.  The invariant mass of the other lepton pair is required to be $40\gev <m_{34}< 115 \gev$.  The dominant SM backgrounds are $ZZ^*$, $Z$, $Zb\bar{b}$, and $t\bar{t}$ and the total background at 4.8 fb$^{-1}$ is $9.3\pm1.2$ events (see Table 3 in~\cite{ATLAS-CONF-2011-162}).  

Parton level signal events were generated in \texttt{MadGraph 4.4.49}~\cite{Alwall:2007st} with CTEQ6L1 parton distribution functions~\cite{Pumplin:2002vw}.  \texttt{Pythia 6.4}~\cite{Sjostrand:2006za} was used for parton showering, hadronization, and particle decays.  The next-to-leading-order (NLO) production cross sections for the electroweakinos were calculated with \texttt{Prospino 2.0}~\cite{Beenakker:1999xh}.  \texttt{PGS}~\cite{PGS} was used as a detector simulator, with a cone jet algorithm with $\Delta R =0.7$.  

To estimate lepton efficiencies, we simulate $h\rightarrow 2l 2\nu$ and $h\rightarrow4l$ in \texttt{MadGraph} for a 150 GeV Higgs and, after applying the cuts described above, compare our results to the published signal distributions and event counts in~\cite{:2011vv, ATLAS-CONF-2011-162}.  For the 4-lepton search, we find an effective muon efficiency of 1 and an effective electron efficiency of 0.85.  For the 2-lepton search, we find an efficiency of 0.6 for electrons with $p_T \lesssim 40 \gev$, rising up to 1 for electrons with $p_T \gtrsim 80 \gev$.  Muons with $p_T \lesssim 35 \gev$ have an efficiency of 0.9, rising up to 1 for muons with $p_T \gtrsim 50 \gev$.  With these lepton efficiencies taken into account, we reproduce the ATLAS distributions to within $\OO{(20\%)}$.   

Next, we study the simplified model introduced in Sec.~\ref{sec: Simplified Model} in the context of these Higgs searches.  We consider the scenario where the $C_1^\pm$ decays entirely to equal combinations of $e^\pm \nu_e N_0^0$ and $\mu^\pm \nu_\mu N_0^0$ or entirely via $W^\pm$.  The former is referred to as ``lepton-dominated'' decays and the latter as ``W-dominated'' decays.  Similarly, the $N_1^0$ can decay to equal combinations of $e^\pm e^\mp N_0^0$ and $\mu^\pm \mu^\mp N_0^0$.  Because the leptonic branching ratios of $Z^0$ decays are so small, no appreciable constraints can be set when these decays dominate. 
\hskip 0.1in
\begin{figure*}[t] 
\includegraphics[width=0.45\textwidth,angle=0]{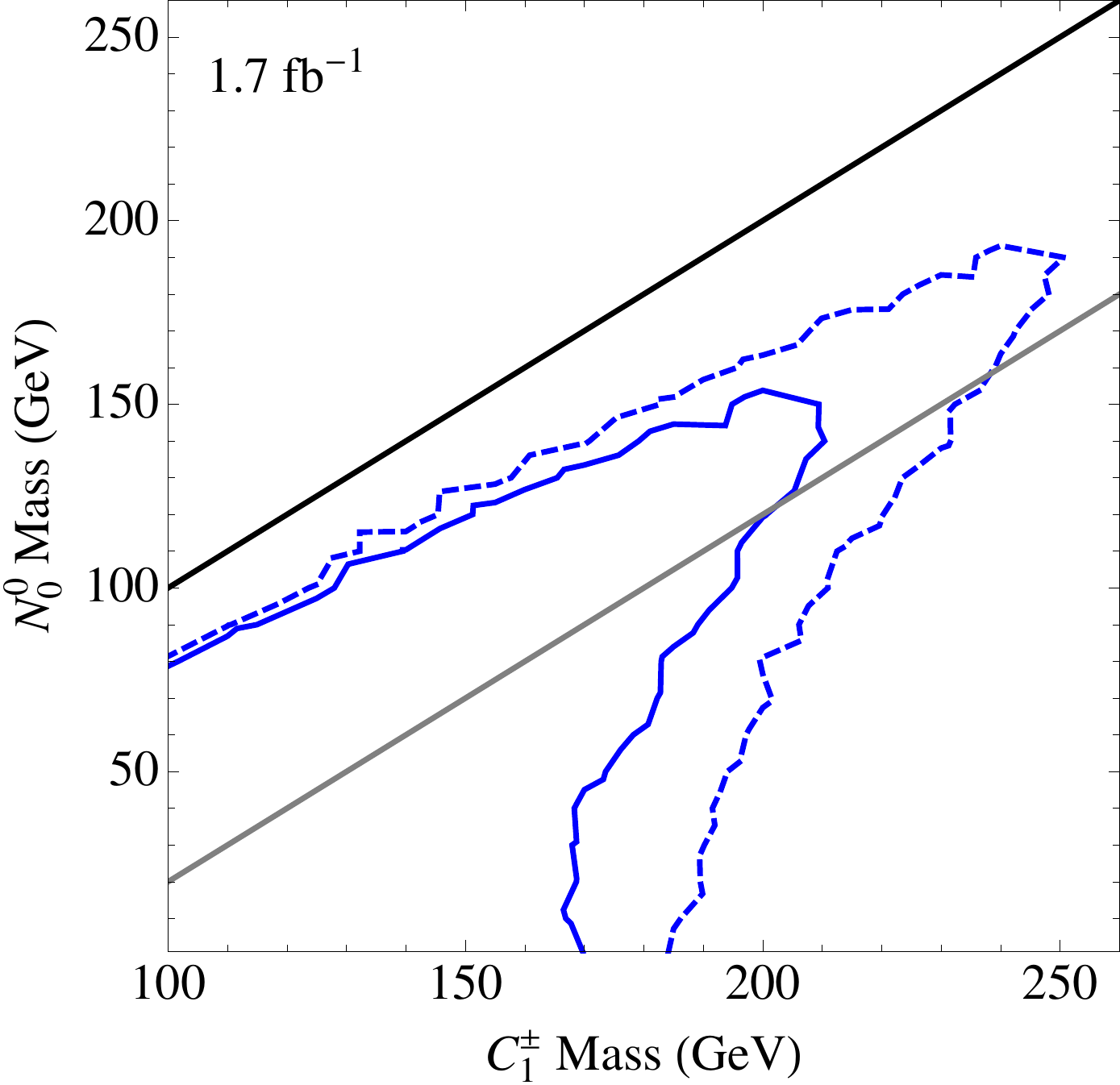}  \hskip 0.2in
\includegraphics[width=0.45\textwidth,angle=0]{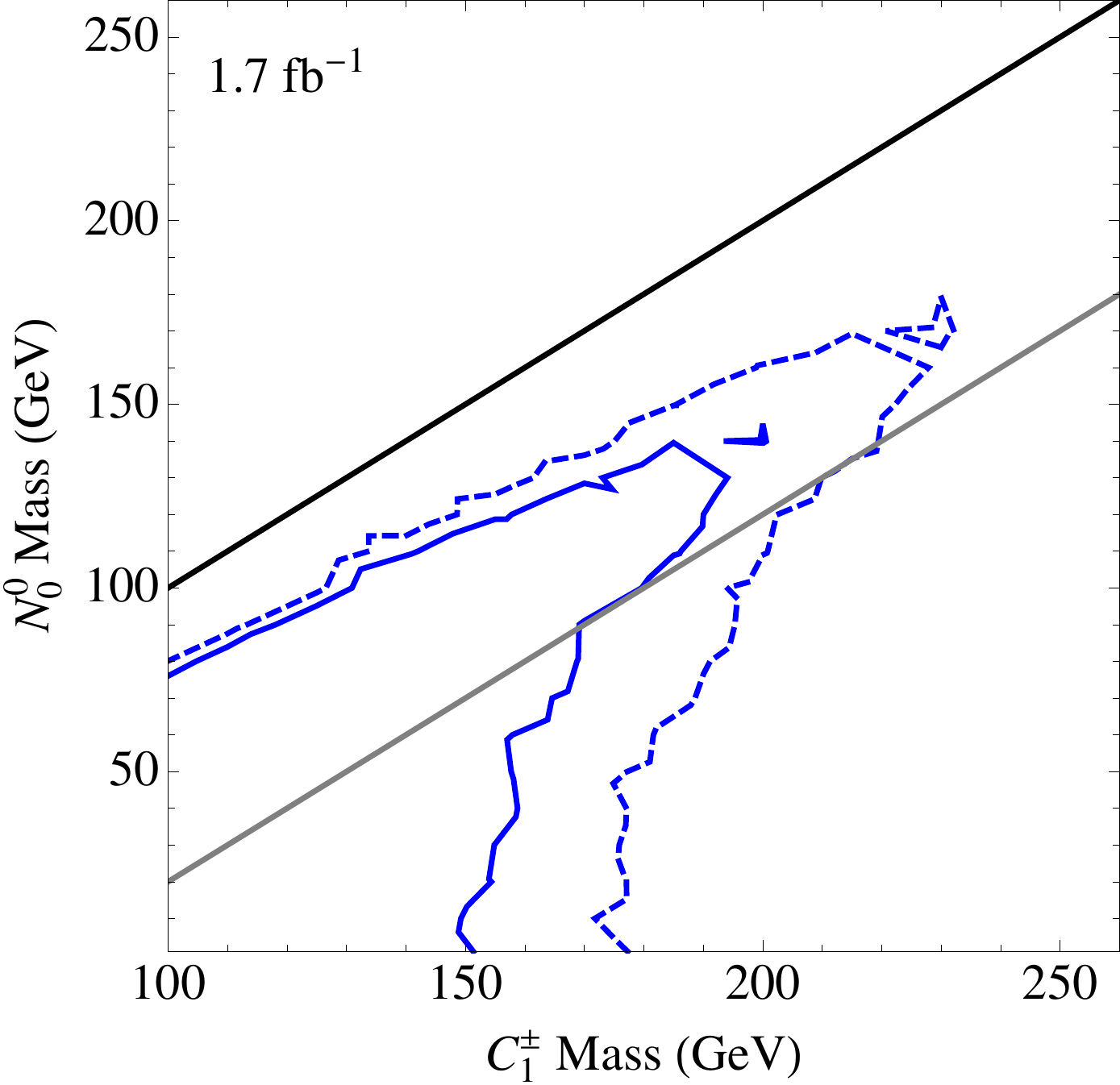} 
  \caption{Constraints from the ATLAS 2-lepton, 0-jet Higgs search on $C^\pm_ 1 N_1^0$ production with lepton and W-dominated decays of charginos (left and right, respectively).  $N^0_1$ is assumed to decay leptonically, and is degenerate in mass with the chargino.  Lines as in Fig.~\ref{fig:charginos}.  }
   \label{fig:2l0j}
\end{figure*}

\subsection{$C_1^{\pm}C_1^{\mp}$ Production}
Figure~\ref{fig:charginos} shows the exclusion limit on the masses of $C_1^{\pm}$ and $N_0^0$ from $C_1^\pm C_1^{\mp}$ pair-production with lepton-dominated decays.  A signal point is taken to be within the 2$\sigma$ sensitivity of the search if the number of signal events is greater than or equal to twice the uncertainty in the background (including both the systematic, $B_{\text{sys}}$, and the statistical, $B_{\text{stat}}$, contributions): 
\begin{equation}
N_{\text{signal}} \ge  2 \times \sqrt{B_{\text{stat}}^2 + B_{\text{sys}}^2}.
\end{equation}
Similarly, the 1$\sigma$ region contains points where the number of signal events is greater than or equal to the background uncertainty.  We use the statistical and systematic uncertainties published by the ATLAS collaboration for their backgrounds. 

As can be seen from Fig.~\ref{fig:charginos}, the 1.7 fb$^{-1}$ 2-lepton, 0-jet Higgs analysis sets 2$\sigma$ limits on the mass of $C_1^{\pm}$ up to $\sim 170 \gev$, for most values of $N_0^0$.  Interestingly, even though the Higgs $WW^*$ analysis is not optimized to look for these electroweak states, it already sets tighter bounds than LEP.  The $2l$+$1j$ Higgs search generally has a lower reach, but provides the possibility of confirming a potential signal.  We have also checked the bounds set by the 8 fb$^{-1}$$WW^*$ Higgs searches at the Tevatron~\cite{DZERO-6219, CDF-10599}.  The Tevatron bounds are not comparable to those set by the LHC, even when assuming that PGS reproduces the Tevatron analysis with 100\% lepton efficiency.   

\subsection{$C_1^{\pm} N_1^0$ Production}
Figure~\ref{fig:2l0j} shows the constraints set by the 2-lepton, 0-jet search on $C_1^\pm N_1^0$ production, for both lepton- and W-dominated decay modes of $C_1^\pm$.  For these plots, $N_1^0$ decays leptonically;  the $Z^0$ decay mode produces too few leptons to set significant limits.  This production mode with lepton-dominated decays typically results in final states with three leptons.  However, the fact that some leptons are lost down the beam pipe or misidentified still allows interesting constraints to be set by the Higgs dilepton analysis.  As a consequence, W-dominated $C_1^\pm$ modes actually have a comparable reach, in spite of having fewer leptons. Additionally, we incorporate the efficiencies by scaling the accepted events by an additional efficiency factor, which is less than one. Some events that are not accepted in our simulation (namely those where all three leptons are detected) could be accepted in practice if one of the leptons were misidentified, not triggering the veto on additional leptons. Thus, our implementation of the efficiencies should be conservative. 

The Tevatron 5.8 fb$^{-1}$ trilepton analysis~\cite{CDFtrileptons} sets a bound of $168 \gev$ on the chargino mass, but only in the mSUGRA scenario.  Figure~\ref{fig:2l0j} shows that the LHC should already be able to exceed this bound in certain regions of parameter space for the simplified model considered here.  In addition, as the LHC Higgs searches accumulate more data, their reach will only continue to improve.

The limits set by the Higgs analysis are stronger than those that would be set by the ATLAS 1 fb$^{-1}$ same-sign dilepton analysis~\cite{Aad:2011cw}.  The same-sign analysis requires that the missing energy in the event be larger than $100 \gev$.  Applying these cuts to the simplified model in this paper and assuming the same lepton efficiencies that we found for the Higgs 2-lepton search, we find that the same-sign search only bounds $C_1^{\pm}$ that are $\gtrsim 80 -140 \gev$ heavier than $N_0^0$, for lepton-dominated decays.  For W-dominated decays, which are hampered by the leptonic branching fraction of the $W^\pm$, the same-sign dilepton search provides no significant limits.  This cross-check confirms our intuition that a large missing energy cut reduces the sensitivity to compressed electroweakino spectra.  

\subsection{$N_1^0 N_1^0$ Production}
\begin{figure}[t] 
\includegraphics[width=0.45\textwidth,angle=0]{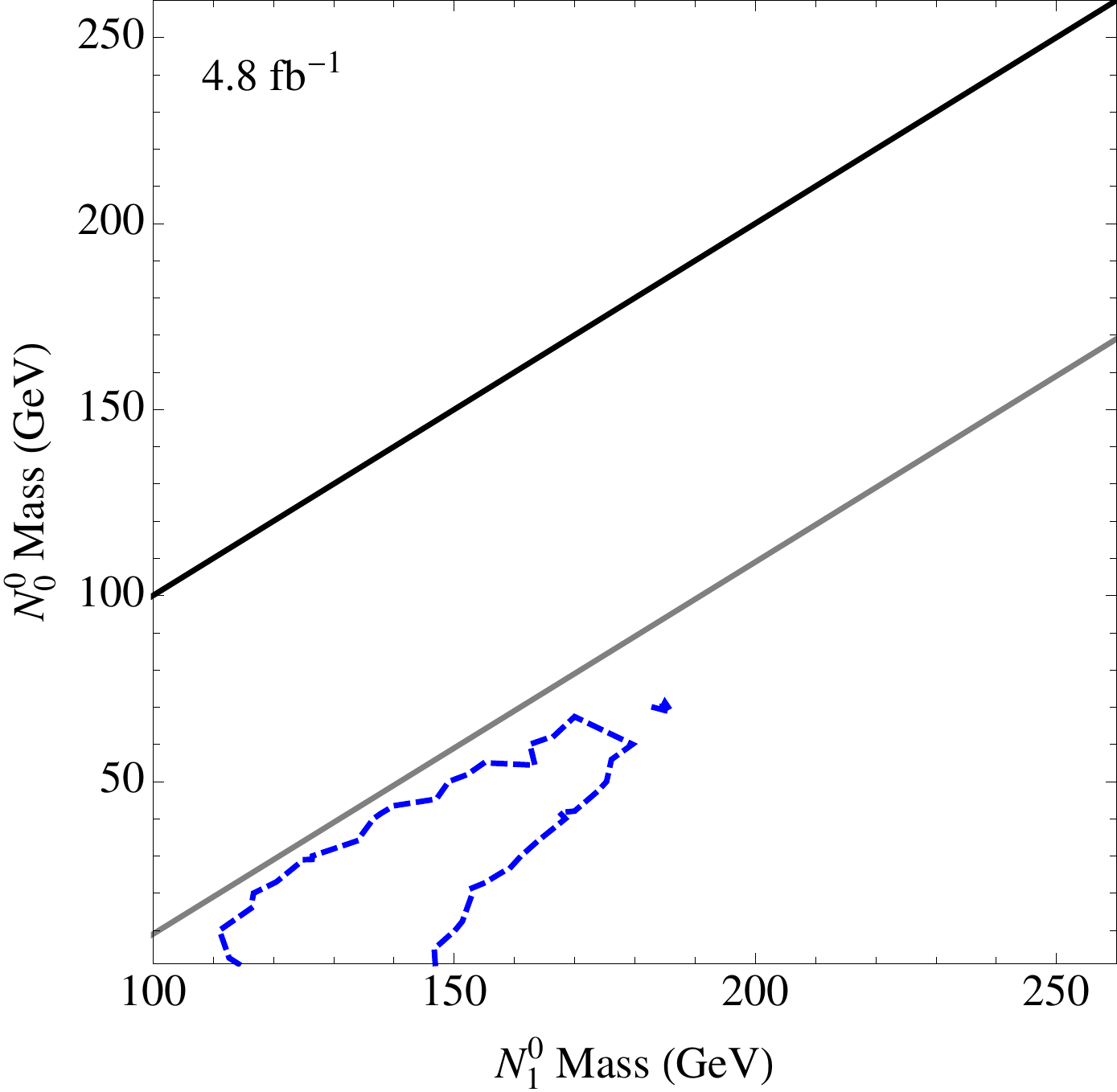}   
 \caption{Sensitivity of the $4$-lepton Higgs search on $N_1^0 N_1^0$ production, with lepton-dominated decays assumed. The production cross section is taken with 800 \gev\, squarks.  Lines as in Fig.~\ref{fig:charginos}, except that the region below the gray line is where on-shell $Z$-boson decays are kinematically allowed.}
   \label{fig:4l}
\end{figure}

The dominant production modes for the 2-lepton electroweakino searches arise from s-channel gauge boson exchange and only have model dependence in the decays.  In comparison, the production mode of $N_1^0 N_1^0$, which contributes to the 4-lepton search, relies on the presence of a higher-dimension production operator (i.e., squark exchange).  Here, we assume $800 \gev$ squarks, which are consistent with current bounds on squark masses from the 1 fb$^{-1}$ ATLAS jets plus missing energy search~\cite{ATLAS-CONF-2011-155}.  Figure~\ref{fig:4l} shows the constraints set by the 4-lepton Higgs search on $N_1^0N_1^0$ production.  Sensitivity is only present in the region where on-shell decays of the Z-boson are allowed.  This is due to the fact that the ATLAS Higgs analysis requires a pair of leptons to reconstruct to the mass of the $Z^0$, which limits the sensitivity of the search in the region of more compressed electroweakino spectra.  Relaxing this requirement (or providing the count rates before applying this cut) could potentially increase sensitivity in this region.
\hskip 0.1in

\subsection{Summary}
As illustrated by Fig.~\ref{fig:charginos}--\ref{fig:4l}, the Higgs $WW^*$ and $ZZ^*$ analyses can set significant limits on the masses of electroweakino states.  It is perhaps striking that an opposite-sign dilepton search can have this reach. The simple fact that the cross section for electroweakino production is in general larger than Higgs production cross sections (for which we know the LHC has sensitivity) promises us some reach. On the other hand, since the dominant background in the 0- and 1-jet searches is WW, making a similar search using same-sign leptons would likely cut backgrounds, but would also reduce signal, if for no other reason but combinatorics. 

We emphasize that the Higgs searches are merely complementary to existing mutlilepton searches, providing sensitivity in the light/compressed regions of parameter space.  The sensitivity of {\em existing} Higgs searches to new electroweakino states motivates further exploration of low-$\MET$ mutlilepton signatures.
\section{Discussion}
\label{sec: Conclusions}

The search for the Higgs boson provides the prospect of discovering the one remaining, untested element of the Standard Model---the origin of electroweak symmetry breaking. However, these same searches are capable of probing ranges of parameters for other electroweakly charged particles.

While many searches for new physics focus on hard jets or large missing energy, the Higgs searches are fairly inclusive for new states in the 100--200 \gev\, mass range. We have considered these searches in the context of electoweakino production as represented by a simplified model. We find that for certain ranges of parameters, Higgs searches are already probing new ground for these particles, placing stronger limits than those from the Tevatron. 

We should emphasize that these limits are not generic, and rely upon certain electroweakino decays dominating. In particular, while interesting limits can arise when the chargino decays through a $W^\pm$,  most limits disappear when $Z^0$ bosons dominate the decays.  Both of these decays are mediated by higher dimension operators, and thus the relative sizes of these modes depend on additional physics, such as the masses of the Higgsinos and sleptons.  Nonetheless, it is important to recognize that new parameter space is already being probed by Higgs searches.  Moreover, the opposite-sign dilepton searches are sensitive to new models (such as Dirac gauginos in e.g., \cite{Hall:1990hq,Fox:2002bu,Kribs:2007ac,Benakli:2011vb,Frugiuele:2011mh,Kribs:2010md}) that same-sign searches may not be sensitive to.

At the same time, it is interesting to note that these new states can form genuine physics backgrounds for Higgs searches. While it is unlikely that an excess of events would be long confused with a Higgs signal (with the more definitive mass peaks from $\gamma \gamma$ and 4-lepton searches being clear cross-checks), a relative failure compared to expectations in one of these channels could indicate the presence of a new, electroweakly charged state. 

As the LHC moves to larger luminosities, the sensitivity to these new states will only increase. This gives the exciting prospect that Higgs searches may not only provide the final piece of the Standard Model, but also the first piece of what comes next.

\vskip0.1in
\noindent{\bf Note added:} As this work was being completed, a related work \cite{ContrerasCampana:2011aa} appeared. There, a complementary study was performed, showing how multilepton searches (optimized for new physics) can be sensitive to Higgs decays.

\section*{Acknowledgements}
We thank Nima Arkani-Hamed, Nathaniel Craig, Patrick Fox, and Natalia Toro for useful conversations.  NW is supported by NSF grant \#0947827. ML acknowledges support from the Simons Postdoctoral Fellows Program.  This work was supported in part by the U.S. National Science Foundation, grant NSF-PHY-0705682, the LHC Theory Initiative, Jonathan Bagger, PI.

\bibliography{multileptons}
\bibliographystyle{apsrev}

\end{document}